\documentclass{rmf-d}
\usepackage{nopageno,rmfbib,multicol,times,epsf,amsmath,amssymb,cite}
\usepackage[latin1]{inputenc}
\usepackage[]{caption2}
\usepackage{graphics}
\usepackage{graphicx}
\def\rmfcornisa{Research in Physics \hfill\rmf\ {\bf } () 
\hfill}

\def\rmfcintilla{{\it Rev.\ Mex.\ Fis.\/} {\bf} () () }
\clearpage \rmfcaptionstyle 
\begin{document}
\title{The odderon discovery by the D0 and TOTEM collaborations
\vspace{-6pt}}
\author{Christophe Royon }
\address{The University of Kansas, Lawrence, USA}
\maketitle

\recibido{20 January 2022}{ \vspace{-12pt} }

\begin{abstract}
\vspace{1em} We describe the discovery of the colorless $C$-odd gluonic compound, the odderon, by the D0 and TOTEM Collaborations by comparing elastic differential cross sections measured in $pp$ and $p \bar{p}$ interactions at high energies.
   \vspace{1em}
\end{abstract}
\begin{multicols}{2}

\section{Introduction: $pp$ and $p \bar{p}$ scattering and the odderon}

\subsection{$pp$ and $p \bar{p}$ scattering}

In this short report, we will study elastic interactions at high energies in order to obtain evidence for the existence of the odderon~\cite{ourpaper,press}.
Elastic $pp$ and $p \bar{p}$ scattering that we are going to study corresponds to the $pp \rightarrow pp$ and $p \bar{p} \rightarrow p \bar{p}$ interactions where the protons and antiprotons are intact after interaction and scattered at very small angle, and nothing else is produced. In order to measure these events, it is necessary to detect the intact protons/antiprotons after interactions in dedicated detectors called roman pots and to veto on any additional activity in the main detector. The fact that the protons (or antiprotons) are intact in the final state means that there is no color exchange between the protons, or in terms of QCD, there must be a two, three, four, five, etc., gluon exchange but not a single one. 
In order to measure elastic interactions, it is needed to detect the intact protons in the final state using roman pot detectors. These detectors can move very close to the beam (up to 3$\sigma$ at the LHC) when beams are stable.

As we mentioned already, elastic scattering is due to the exchange of colorless objects (pomeron and odderon).
Pomeron and odderon correspond to positive and negative charge ($C$) parity.  The odderon is defined as a singularity in the complex plane, located at $J = 1$ when $t = 0$ and which contributes to the odd crossing amplitude~\cite{nicolescu,martynov,landshoff,leader}. From the point of view of QCD, the pomeron is made of an even number of  gluons (two, four, etc...) which leads to a ($+1$) parity whereas the odderon is made of an odd number of gluons (three, five, etc...) corresponding to a ($-1$) parity. The scattering amplitudes can be written as the sum or the difference of the even and the odd part of the amplitude for $pp$ and $p \bar{p}$ scattering
\begin{eqnarray}
A_{pp} &=& Even~+~Odd \nonumber \\
A_{p \bar{p}} &=& Even~-~Odd. \nonumber
\end{eqnarray}
From the equations above, it is clear that observing a difference between $pp$ and $p \bar{p}$ interactions could be a clear signal for the odderon.

\subsection{Generic behavior of elastic $d \sigma/dt$ cross section}

\begin{figure*}
\centering
\includegraphics[width=0.6\textwidth]{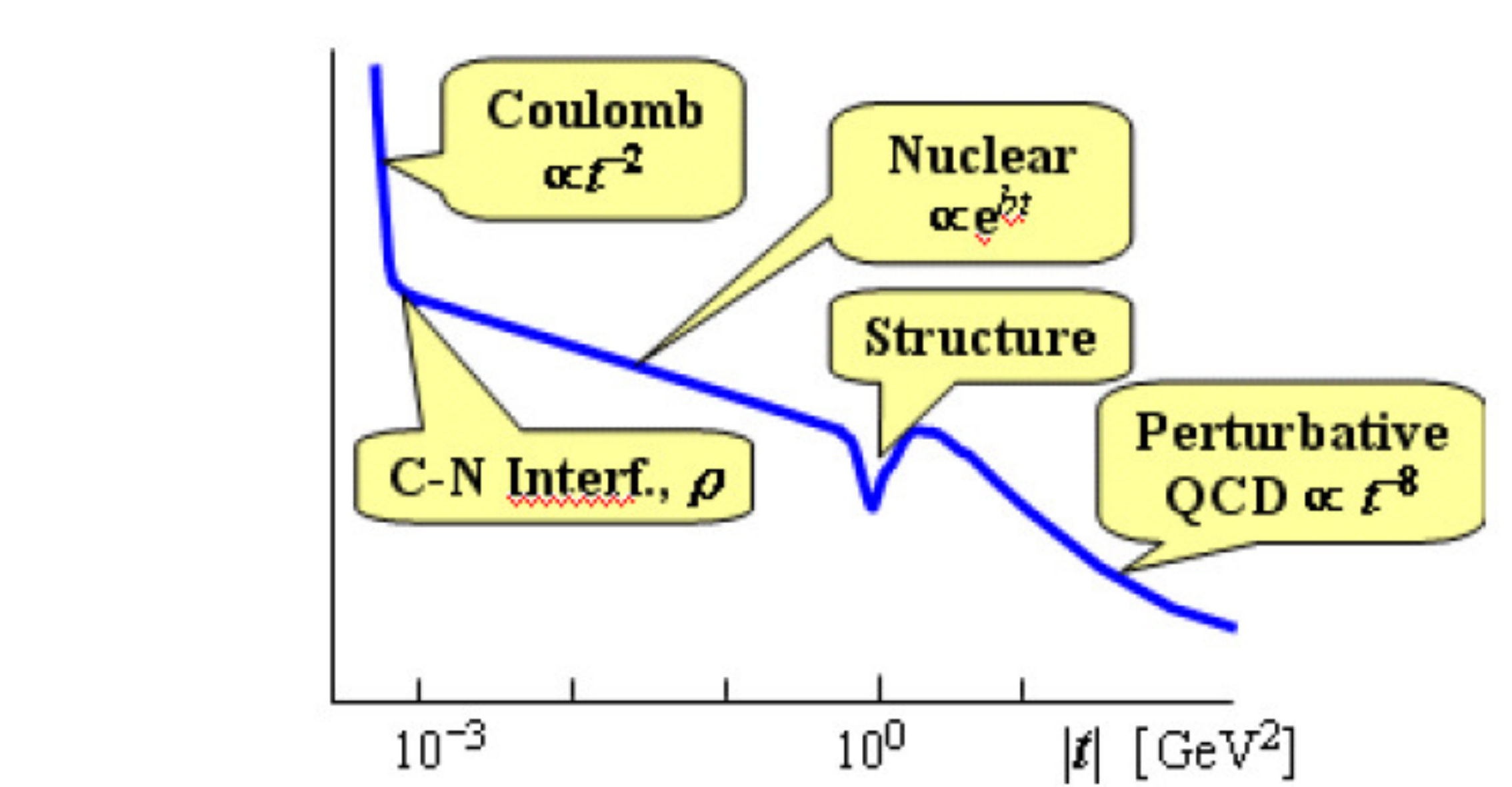}
\caption{Schematic of elastic $d \sigma/dt$ cross section as a function of $t$.}
\label{fig0}
\end{figure*}

Before describing the measurements of elastic cross sections at high energy from the Tevatron and the LHC, let us describe briefly the generic behavior of the elastic $d \sigma /dt$ cross section as a function of $|t|$ as shown in Fig.~\ref{fig0}.  Measuring $t$ is performed via tracking the intact protons through the machine, using the magnets as a spectrometer (since the protons lose part of their momentum, they are scattered away from the beam). At very low $|t|$ ($|t| \sim 10^{-4}$, $10^{-3}$ GeV$^2$), we are in the Coulomb QED region, and the cross section decreases as $|t|^{-2}$. At higher $|t|$, we reach the nuclear region where the cross section decreases exponentially (between the two regions, there is the Coulomb-nuclear interference region which is fundamental for the odderon discovery as we will see in the following). At high $|t|$ ($|t| \sim 1$ GeV$^2$), we can observe some structure (maxima and minima called respectively bumps and dips), and finally at even higher $|t|$ ($|t| > 3-4$ GeV$^2$), we reach the perturbative QCD region.

\begin{figure*}
\centering
\includegraphics[width=0.4\textwidth]{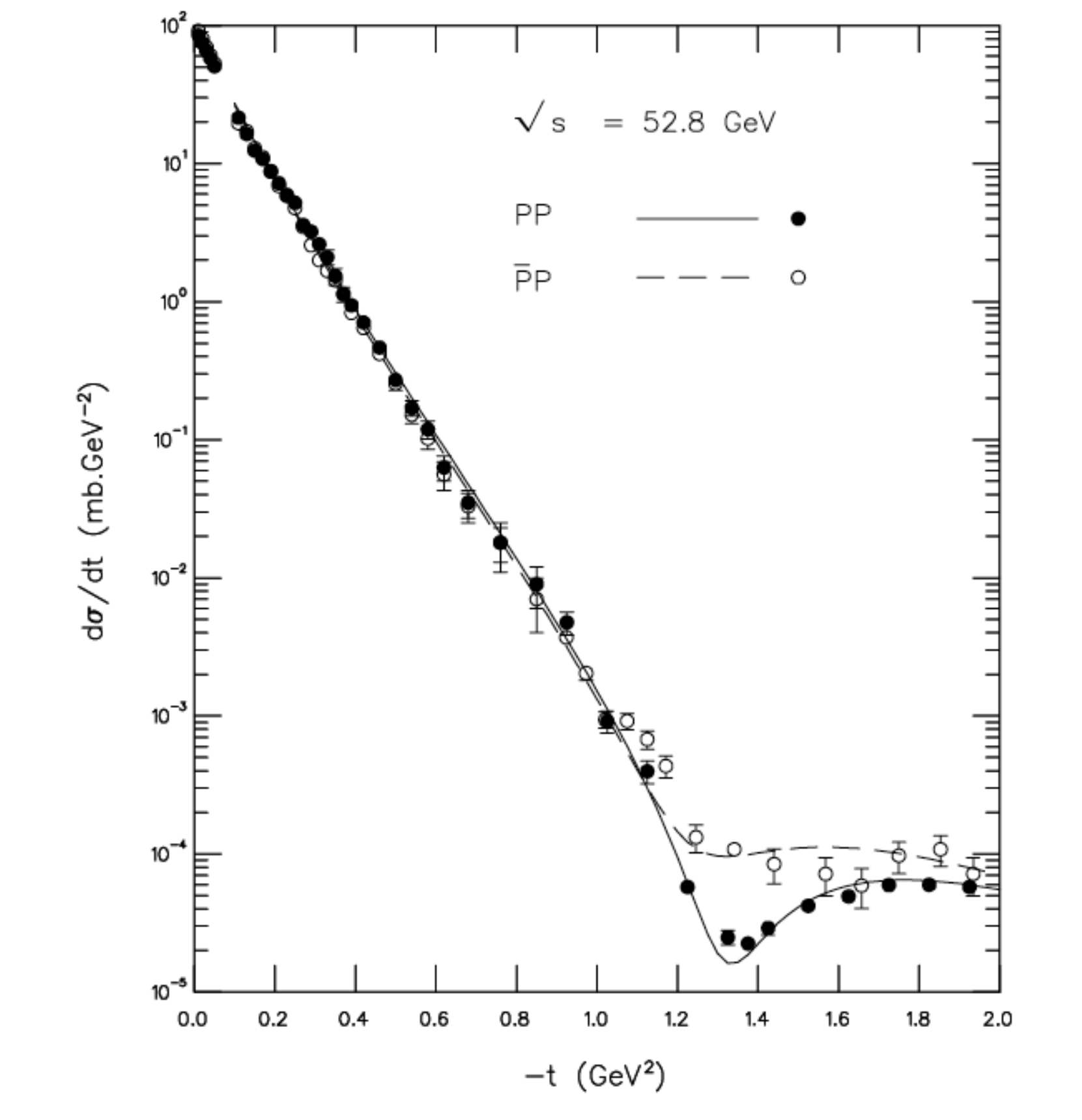}
\caption{Measurement of elastic $pp$ and $p \bar{p}$ $d\sigma/dt$ at ISR energies.}
\label{fig0b}
\end{figure*}

Many experiments have been looking for evidence of the existence of the odderon in the last 50 years, and one may wonder why the odderon has been so elusive. At ISR energies, at about a center-of-mass energy of 52.8 GeV~\cite{ISR}, there was already some indication of a possible difference between $pp$ and $p\bar{p}$ interactions as shown in Fig.~\ref{fig0b}. Differences are about 3$\sigma$ but this was not considered to be a clean proof of the odderon.  This is due to the fact that elastic scattering at low energies can be due to exchanges of additional particles to pomeron and odderon, namely $\rho$, $\omega$, $\phi$ mesons and reggeons. It is not easy
to distinguish between all these possible exchanges, and it becomes quickly model dependent. This is why the observed difference at 52.8 GeV was estimated to be due to $\omega$ exchanges and not to the existence of the odderon.

The advantage of being at higher energies (1.96 TeV for the Tevatron and 2.76, 7, 8 and 13 TeV at the LHC) is that meson and reggeon exchanges can be neglected (this is shown by the smoothness of the $|t|$ dependence of the elastic $d \sigma/dt$ cross section as measured by the TOTEM collaboration at 7, 8 and 13 TeV~\cite{totemdata} that do not show any dips or bumps at higher values of $|t|$). It means that a possible observation of differences between $pp$ and $p \bar{p}$ elastic interactions at high energies would be a clear signal of the odderon.


\begin{figure*}
\centering
\includegraphics[width=0.45\textwidth]{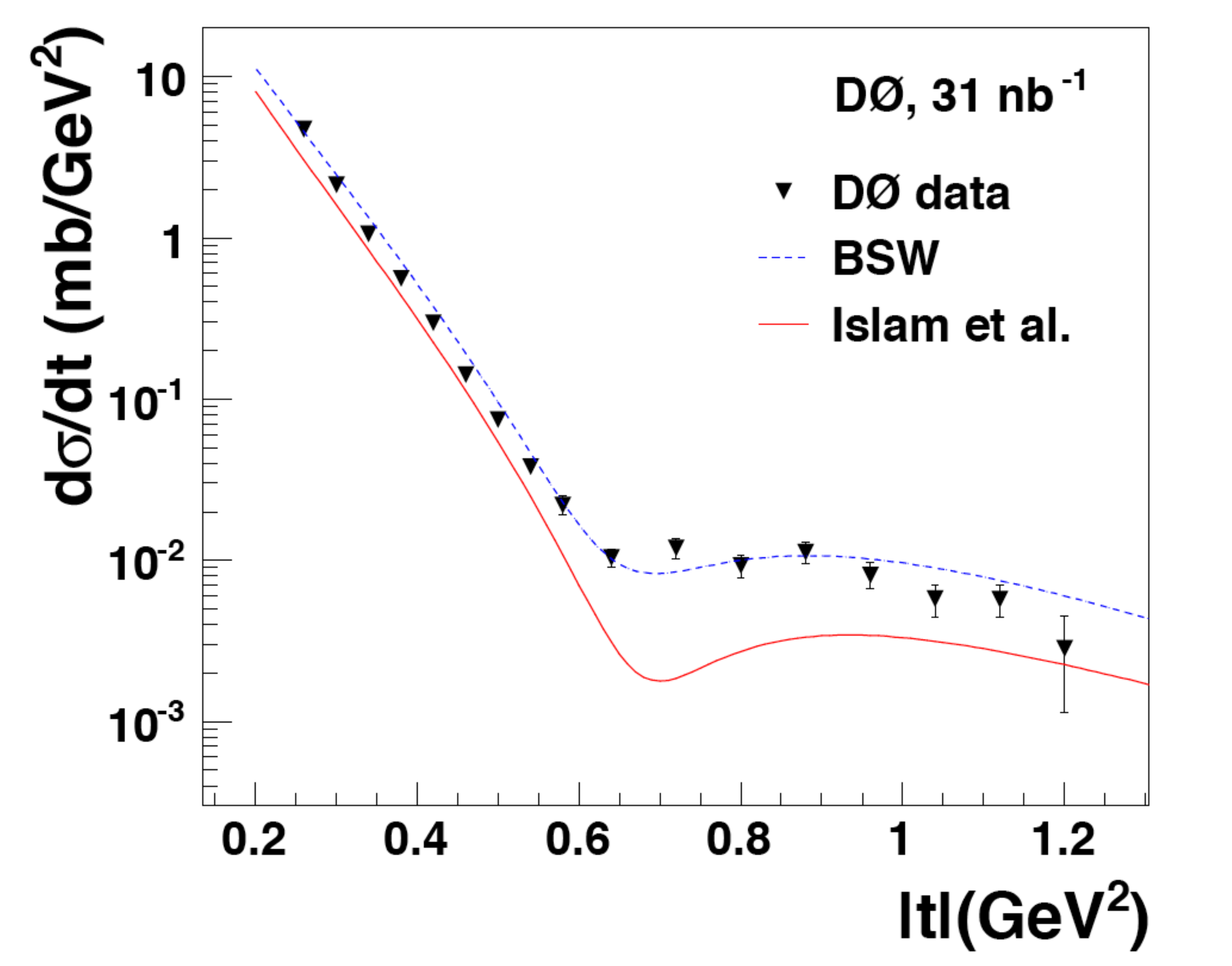}
\includegraphics[width=0.5\textwidth]{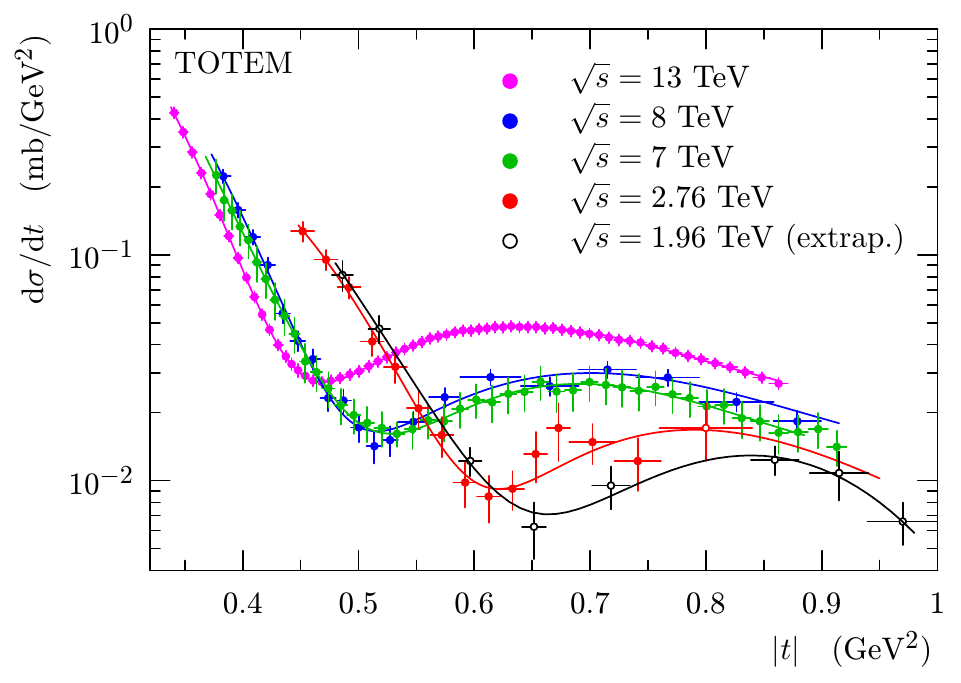}
\caption{Left: $p \bar{p}$ elastic cross section as a function of $|t|$ at 1.96 TeV from the D0 collaboration at the Tevatron. Right: $pp$ elastic cross sections as a function of $|t|$ at 2.76, 7, 8, and 13 TeV from the TOTEM collaboration at the LHC (full circles), and extrapolation to the Tevatron center-of-mass energy at 1.96 TeV (empty circles).}
\label{fig1}
\end{figure*}

\section{Strategy to compare $pp$ and $p \bar{p}$ elastic scattering and measurement of the ratio of the elastic $d \sigma/dt$ cross section at the bump and at the dip}
In this section, we will detail the method to compare elastic scattering at high energy between $pp$ and $p \bar{p}$ interactions. The D0 collaboration installed roman pot detectors on both sides of the main D0 detector~\cite{FPD} to measure intact $p$ and $\bar{p}$ in the final states after interaction at a center-of-mass energy of 1.96 TeV.  Using  31 nb$^{-1}$ of data, the D0 collaboration measured the elastic $p\bar{p}$  $d\sigma/dt$ at 1.96 TeV for $0.26<|t|<1.2$ GeV$^2$ ~\cite{d0cross} as shown in Fig.~\ref{fig1}, left. 
The TOTEM collaboration at the LHC installed some roman pot detectors on both sides of the CMS detectors (interaction point number 5 at the LHC) at about 220 m from the interaction point in order to measure intact protons. This allows measuring the $pp \rightarrow pp$ elastic cross section at center-of-mass energies of 1.96, 7, 8 and 13 TeV by vetoing on any activity in the main CMS detector as shown in Fig.~\ref{fig1}, right. In addition, the TOTEM collaboration installed two telescopes covering a domain in rapidity  $3.1 < |\eta| < 4.7$ and $5.3< |\eta| < 6.5$, allowing to veto on particle production at very large rapidities in the very forward region. The advantage of the LHC is that it is possible to run the machine at different center-of-mass energies and also with different values of $\beta^*$~\footnote{$\beta^*$ describes how the beams are parallel between each other.} that allow to cover a large domain in $|t|$ for elastic cross section measurements. As an example, at 13 TeV, a value of $\beta^*$ of 2.5 km (respectively 90 m) allows covering a domain $3. 10^{-4} < |t| < 0.2$ GeV$^2$ (respectively $5. 10^{-3} < |t| < 3.5$ GeV$^2$). These two aspects will be fundamental for the odderon discovery. 

In order to understand the features of $pp$ elastic scattering interactions at high energies, the measurements of the $pp$ elastic $d \sigma/dt$ cross sections by the TOTEM collaboration at different center-of-mass energies in the same $|t|$ domain as the measurement by the D0 collaboration for $p \bar{p}$ interactions are shown in Fig.~\ref{fig1}, right. Data always show the same features namely a decrease of the elastic cross section at lower $|t|$, then the presence of a dip, an increase at higher $|t|$, the presence of a bump, and finally a decrease at higher $|t|$. The D0 $p \bar{p}$ measurement shown in Fig.~\ref{fig1}, left, does not show the same feature, the cross section decreases at lower $|t|$, reaches a plateau, and decreases (within uncertainties) at higher $|t|$. There is no dip and bump observed in $p \bar{p}$ interactions. This observation will be the origin of the method to measure quantitatively the differences between $pp$ and $p \bar{p}$ elastic interactions~\cite{ourpaper}. It is clear that this comparison can only be performed in the common domain in $|t|$ between the D0 and TOTEM measurements in oder to avoid any extrapolation.

\begin{figure*}
\centering
\includegraphics[width=1.0\textwidth]{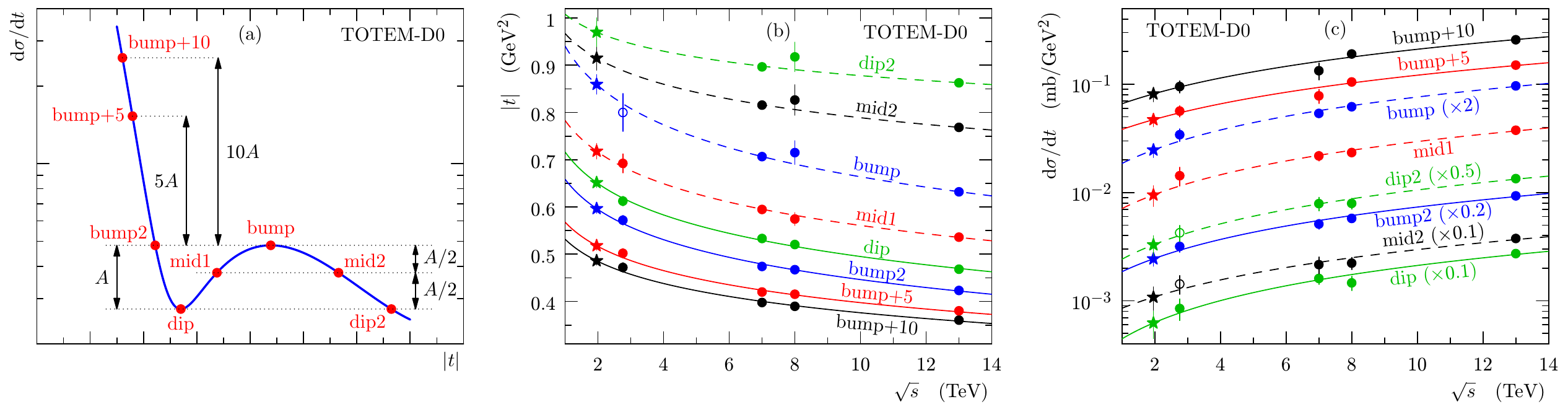}
\caption{$(a)$ Schematic definition of the reference points in the TOTEM elastic differential cross section data.  $(b)$ and $(c)$ $|t|$ and $d\sigma/dt$ values for all reference points from TOTEM measurements at 2.76, 7, 8, and 13 TeV (circles) as a function of $\sqrt{s}$ extrapolated to the Tevatron center-of-mass energy (stars).  }
\label{fig3}
\end{figure*}

In order to quantify the differences, we chose to define 8 reference points that are characteristic of the behavior of elastic $pp$ $d \sigma /dt$ as illustrated in Fig.~\ref{fig3}, right. We measure both the $|t|$ and $d \sigma/dt$ values for each characteristic point. The two first points correspond to the $dip$ and the $bump$. Additionally, we define $dip2$, $bump2$ (same values of $d \sigma/dt$ as at the dip and the bump but respectively at higher and lower $|t|$), $mid1$, $mid2$ (middle in $d \sigma/dt$ between the bump and the dip), $bump+5$, $bump+10$ ($d\sigma/dt$ at lower $|t|$ that corresponds to 5 and 10 times the difference in cross section between the bump and the dip). The fact to choose 8 points is of course somewhat arbitrary. We use data points closest to those characteristic  points in order to avoid model-dependent fits. 

\begin{figure*}
\centering
\includegraphics[width=0.45\textwidth]{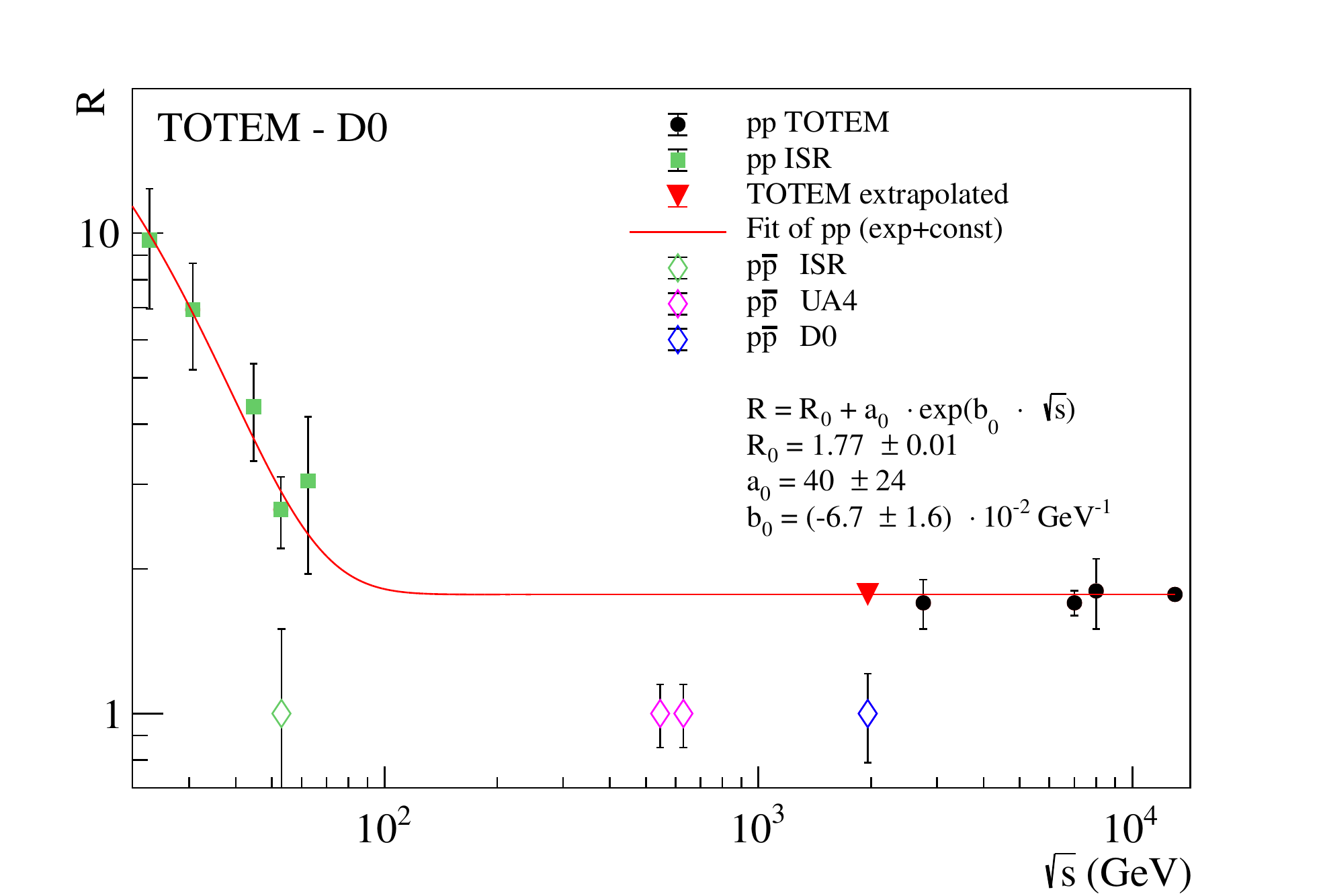}
\caption{$R$ Ratio of the elastic $d\sigma/dt$ cross sections at the bump and the dip as a function of $\sqrt{s}$ compared between $pp$ and $p \bar{p}$ elastic
interactions}
\label{fig2}
\end{figure*}

The first simplest observable is the ratio $R$ of the elastic $d \sigma/dt$ cross section at the bump and at the dip. The ratio $R$ is shown as a function of $\sqrt{s}$ in Fig.~\ref{fig2}. $R$ is displayed at ISR~\cite{ISR} and LHC energies for $pp$ interactions respectively in green and black full points.
$R$ decreases as a function of $\sqrt{s}$ up to $\sim$100 GeV and is flat above, allowing to extrapolate the TOTEM measurements at 2.76, 7, 8 and 13 TeV to the Tevatron energy of 1.96 TeV.  It is worth noticing the decrease of $R$ at lower energies, which shows the already mentioned different behavior of elastic interactions at low energies due to the additional exchanges of mesons, reggeons that appear in this domain. 
$p \bar{p}$  measurements~\cite{ISR}  show a ratio of 1.00 (with an uncertainty of 0.21 for D0) given the fact that no bump nor dip is observed in $p \bar{p}$ data within uncertainties. It leads to  a difference by more than 3$\sigma$  between $pp$ and $p \bar{p}$ elastic data~\cite{ourpaper} (assuming a flat behavior above $\sqrt{s}$ of about 100 GeV).

The definition of the 8 reference points defined above leads to a distribution of $t$ and $d \sigma/dt$ values as a function of $\sqrt{s}$ for all points as shown in Fig.~\ref{fig3}, middle and right. In order to be able to compare with the D0 $p \bar{p}$ elastic $d \sigma/dt$ cross section measurement, it is needed to extrapolate the $|t|$ and $d \sigma/dt$ values of these reference points to the Tevatron energy of 1.96 TeV.  To do so, we perform a fit of the $|t|$ and $d \sigma/dt$ values of the reference points using the following formulae
\begin{eqnarray}
|t| &=& a \log (\sqrt{s}{\rm [TeV]}) + b  \label{eqn2} \nonumber \\
(d\sigma/dt) &=& c \sqrt{s}~{\rm [TeV]} + d. \label{eqn3} \nonumber 
\end{eqnarray}
The same form is used for the 8 reference points (this is an assumption and works to describe all characteristic points) and this simple form is chosen since we fit at most 4 points, corresponding to $\sqrt{s}= $ 2.76, 7, 8 and 13 TeV.
We also tried alternate parametrizations such as $|t|=e (s)^f$ leading to compatible results well within 1$\sigma$ uncertainty,
leading to very good $\chi^2$ per dof, better than 1. It is also worth noticing that the 2.76 TeV measurements are crucial to reduce the range of extrapolation in energy between the Tevatron and LHC center-of-mass energies. A direct comparison between D0 and TOTEM data (by running the LHC at 1.96 TeV) is unfortunately not possible since it is difficult to run the LHC at this center-of-mass energy and the present location of the TOTEM roman pot detectors does not give any acceptance in the bump/dip domain in $|t|$, making a $pp$ measurement of the elastic cross section impossible at this stage. The only possibility to compare $pp$ and $p \bar{p}$ elastic interactions in the same $t$ region is thus to extrapolate the TOTEM measurements down to the Tevatron energy.

\section{Comparison between the elastic $d\sigma/dt$ measurements from D0 and the extrapolated TOTEM data and the odderon discovery}

\subsection{Fits of TOTEM extrapolated characteristic points at 1.96 TeV}

The last step is to predict the $pp$ elastic cross sections at the same $|t|$ values as measured by D0 in order to make a direct comparison. We thus fit the reference points extrapolated to 1.96 TeV from the TOTEM measurements as a function of $|t|$  using a double exponential fit ($\chi^2=0.63$ per dof)
\begin{eqnarray}
h(t) = a_1 e^{-b_1 |t|^2 - c_1|t|}  
+ d_1 e^{-f_1 |t|^3 - g_1 |t|^2 -h_1 |t|}.
 \nonumber
\end{eqnarray}
This function is chosen for fitting purposes only, the first function describing the low-$t$ diffractive cone and the second one the asymmetric structure at the bump and the dip.
The two exponential terms cross around the dip, one rapidly falling and becoming 
negligible in the high $|t|$-range while the other term rises above the dip.
Systematic uncertainties are evaluated from an ensemble of MC
experiments in which the cross section values of the eight characteristic points are varied within their Gaussian uncertainties. Fits without a dip and bump position matching the extrapolated values within their uncertainties are rejected, and slope and intercept constraints are used to discard unphysical fits. It is also worth noting that such a simple
formula leads also to a good description of TOTEM data in the dip and bump region at 2.76, 7, 8 and 13 TeV.

\begin{figure*}
\centering
\includegraphics[width=0.4\textwidth]{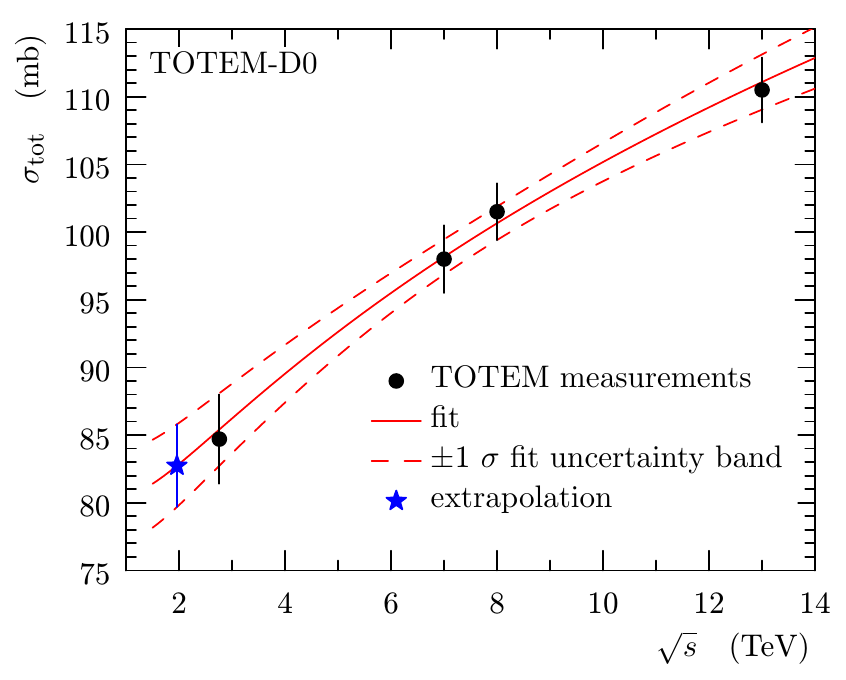}
\caption{Total cross section as a function of $\sqrt{s}$ measured by the TOTEM collaboration (back points) extrapolated to the Tevatron center-of-mass energy (blue star).}
\label{fig4}
\end{figure*}

\subsection{Relative normalization between D0 measurement and extrapolated TOTEM data: total $pp$ cross section at 1.96 TeV}

Since we want to compare the shape of the elastic $pp$ and $p \bar{p}$ $d \sigma/dt$ cross section measurements, we need to adjust the TOTEM and D0 data sets to have the same cross sections at the optical point (OP)
$d\sigma/dt(t = 0)$ (OP cross sections are expected
to be equal if there are only $C$-even exchanges, and we add an additional 3\% systematic uncertainty that would originate of the expected difference in the case of the maximal odderon models). The fully correlated uncertainty due to the D0 luminosity measurement is indeed about 12.5\%. To do so, we use the optical theorem that relates the total cross section $\sigma_{tot}$ to the slope of the elastic cross section at $t=0$
\begin{eqnarray}
\sigma_{tot}^2 = \frac{16 \pi (\hbar c)^2}{1+\rho^2}  \left( \frac{d \sigma}{dt} \right)_{t=0}. \nonumber
\end{eqnarray}

The first step is to predict the $pp$ total  cross section from the extrapolated fit to TOTEM data 
\begin{eqnarray}
\sigma_{tot} = a_2 \log^2 \sqrt{s}{\rm [TeV]} +b_2  \nonumber
\end{eqnarray}
with a $\chi^2$ of 0.27 as illustrated in Fig.~\ref{fig4}.  It is worth noting that this formula is obviously only valid in the region above 1 TeV where the fit is performed and not at lower energies. 
Other parametrizations such as ($A log^2(\sqrt{s}) +B log(\sqrt{s})+C$), ($As + B \sqrt{s} +C$) and ($A+B s^{0.25}$)  lead to similar results. It 
leads to an estimate of the $pp$ total cross section of $\sigma_{tot}=$82.7 $\pm$ 3.1 mb at 1.96 TeV. 
Using the optical theorem and assuming $\rho=0.145$ (the ratio of the imaginary and the real part
of the elastic amplitude at $t=0$), as taken from COMPETE extrapolation~\cite{compete},  leads to a TOTEM $d\sigma/dt(t=0)$ at the OP of 357.1 $ \pm$ 26.4 mb/GeV$^2$. The D0 collaboration measured the optical point of $d \sigma/dt$ at small $t$ to be 341$\pm$48 mb/GeV$^2$, and we thus rescale the TOTEM data  by 0.954 $\pm$ 0.071 (let us note that the TOTEM and D0 measurements are compatible within uncertainties before rescaling).
Of course, we do not claim that we performed a measurement of $d\sigma/dt$ at
the OP at $t = 0$ (it would require additional measurements  closer to $t = 0$ especially at Tevatron energies), but
we use the two extrapolations simply in order to obtain a common and somewhat arbitrary normalization point.

\subsection{Comparison between D0 measurement and extrapolated TOTEM data}

\begin{figure*}
\centering
\includegraphics[width=0.5\textwidth]{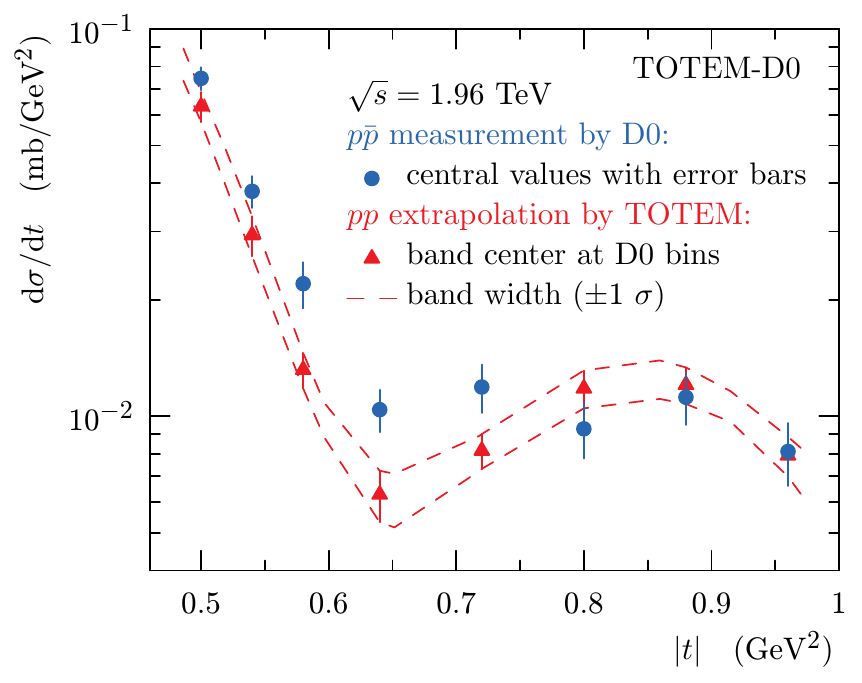}
\caption{Comparison between the D0 $p\bar{p}$ measurement
at 1.96 TeV and the extrapolated TOTEM $pp$ cross section, rescaled to match the OP of
the D0 measurement.  The dashed lines show the
1$\sigma$ uncertainty band on the extrapolated $pp$ cross section.}
\label{fig5}
\end{figure*}

The comparison between the $p \bar{p}$ elastic $d \sigma/dt$ measurement by the D0 collaboration and the extrapolation of the TOTEM $pp$ elastic $d \sigma/dt$ measurements is shown in Fig.~\ref{fig5}, including the 1$\sigma$ uncertainty band as a red dashed line~\cite{ourpaper}. The comparison is only made in the common $t$ domain for both $pp$ and $p \bar{p}$ measurements and show some differences in the dip and bump region between $|t|$ of  0.55 and 0.85 GeV$^2$. The remaining step is to evaluate quantitativaly the difference. We perform a $\chi^2$ test to examine the probability for the D0 and TOTEM $d\sigma/dt$ to agree
\begin{eqnarray}
\chi^2 = \Sigma_{i,j} [(T_i-D_i)C_{ij}^{-1} (T_j-D_j)] + \frac{(A-A_0)^2}{\sigma_A^2} + \frac{(B-B_0)^2}{\sigma_B^2}  \nonumber
\end{eqnarray}
where $T_j$ and $D_j$ are the $j^{th}$ $d\sigma/dt$ values for TOTEM and D0, $C_{ij}$ the covariance matrix, $A$ ($B$) the nuisance parameters  for scale (slope) with $A_0$ ($B_0$) their nominal values. The first constraint $A$ is the matching of the $pp$ and $p \bar{p}$ OP and the second one $B$ is the matching of the $pp$ and $p \bar{p}$ slopes in the diffractive cone region. Potential differences on the slopes could be due to the odderon, but it is known that the pomeron is dominating in the diffractive cone region.
Given the constraints on the OP normalization and logarithmic slopes of the elastic cross sections, the $\chi^2$ test with six degrees of freedom yields the $p$-value of 0.00061, corresponding to a significance of 3.4$\sigma$.

\section{Discovery of the odderon}

\begin{figure*}
\centering
\includegraphics[width=0.7\textwidth]{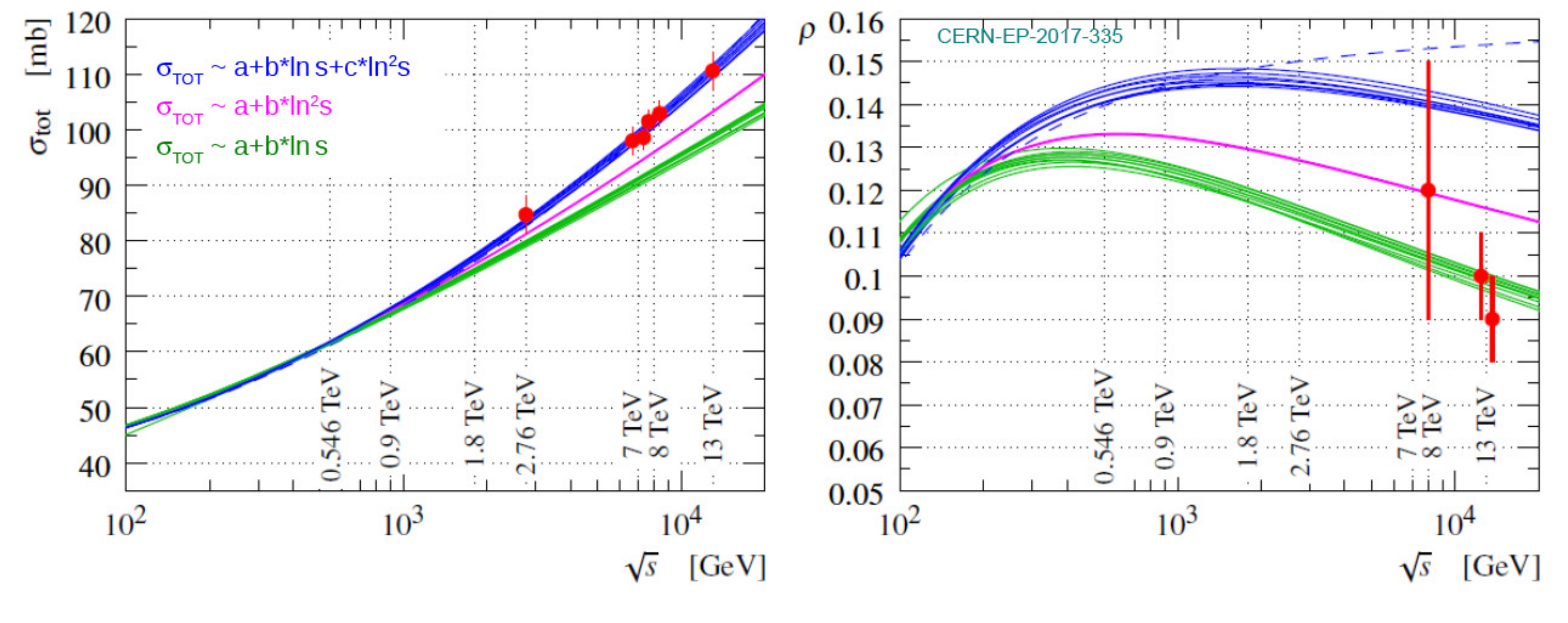}
\caption{Total cross secttion (left) and $\rho$ (right) measurements by the TOTEM collaboration as a function of $\sqrt{s}$ compared to predictions from the COMPETE collaboration.}
\label{fig6}
\end{figure*}

The previous analysis can be combined with a previous measurement performed by the TOTEM collaboration corresponding to the 
measurement of elastic scattering at very low $t$ in the Coulomb-nuclear
interference (CNI) region where
\begin{eqnarray}
\frac{d \sigma}{dt} \sim |A^C + A^N ( 1 -\alpha G(t))|^2. \nonumber
\end{eqnarray}
The differential cross section is sensitive to the phase of the
nuclear amplitude and in the CNI region, both the modulus and the phase of the nuclear 
amplitude can be used to determine 
\begin{eqnarray}
\rho = \frac{Re(A^N(0))}{Im (A^N(0))} \nonumber
\end{eqnarray}
and the total cross section (the modulus is constrained by the measurement in the hadronic region and the
phase by the $t$ dependence).
The measurement of $\rho$ at 13 TeV was $\rho=0.09 \pm 0.01$~\cite{rho}. 
The values of the measured $\rho$ and $\sigma_{tot}$ values are not compatible with any set of models without odderon exchange. Fig.~\ref{fig6} shows that the $\sigma_{tot}$ measurements favor the ($ln s + ln^2 s$) series of parametrizations of COMPETE~\cite{compete} whereas the $\rho$ measurement at 13 TeV favor the ($ln s$) parametrizations.  This tension between these two measurements can be explained by the exchange of the Odderon (none of the COMPETE parametrization contains the odderon). For the models included in 
COMPETE as an example, the TOTEM $\rho$ measurement at 13 TeV provided a 3.4 to 4.6$\sigma$ 
significance. This result is similar for the Durham~\cite{durham} (4.3$\sigma$) and Block-Halzen~\cite{block} (3.9$\sigma$) models.

This result can be combined 
with the D0 and TOTEM result presented in the previous section since it corresponds to completely independent data taking and even detectors (the measurement in the CNI region requires a high values of $\beta^*$ of 2.5 km in order to access very low $|t|$ values whereas the comparison with D0 was performed using data taken at $\beta^*=90$ m for a domain in $|t|$ in the dip and bump region). 
The combined 
significance ranges from 5.3 to  5.7$\sigma$ (depending on the model).
Models without colorless $C$-odd gluonic compound or the odderon are excluded by more than 5$\sigma$.

\section{Conclusion}

We analyzed the differences between elastic $pp$ and $p \bar{p}$ interactions at 1.96 TeV by comparing the measurements of the D0 collaboration and the extrapolation of the TOTEM measurements at 2.76, 7, 8 and 13 TeV.
$pp$ and $p\bar{p}$ cross sections differ
with a significance of 3.4$\sigma$ in a model-independent way and thus provides evidence that the Colorless $C$-odd gluonic compound,  i.e. the odderon,
is needed to explain elastic scattering at high energies.
When combined with the $\rho$ and total cross section result at 13 TeV from the TOTEM Collaboration, the significance is in the range 5.2 to 5.7$\sigma$ and thus constitutes the first experimental observation of the odderon, which represents a  major discovery at CERN and Tevatron~\cite{ourpaper,press}.

\end{multicols}

\medline

\begin{multicols}{2}

\end{multicols}
\end{document}